\documentclass[prd,aps,preprint,nofootinbib,superscriptaddress]{revtex4}
\usepackage{epsfig}
\usepackage{amsmath}

\begin{document}

\preprint{BNL-NT-06/14} \preprint{RBRC-574}

\title{Single Transverse-Spin Asymmetry in Drell-Yan Production \\
at Large and Moderate Transverse Momentum}

\author{Xiangdong Ji}
\email{xji@physics.umd.edu} \affiliation{Physics Department,
University of Maryland, College Park, MD 20742}
\author{Jian-Wei Qiu}
\email{jwq@iastate.edu} \affiliation{Department of Physics and
Astronomy, Iowa State University, Ames, IA 50011}
\author{Werner Vogelsang}
\email{vogelsan@quark.phy.bnl.gov} \affiliation{Physics
Department, Brookhaven National Laboratory, Upton, NY 11973}
\affiliation{RIKEN BNL Research Center, Building 510A, Brookhaven
National Laboratory, Upton, NY 11973}
\author{Feng Yuan}
\email{fyuan@quark.phy.bnl.gov} \affiliation{RIKEN BNL Research
Center, Building 510A, Brookhaven National Laboratory, Upton, NY
11973}
\date{\today}

\begin{abstract}
We study the single-transverse spin asymmetry for the Drell-Yan
process. We consider production of the lepton pair at large
transverse momentum, $q_\perp\sim Q$, where $Q$ is the pair's
mass. The spin asymmetry is then of higher twist and may be
generated by twist-three quark-gluon correlation functions.
Expanding the result for $q_\perp\ll Q$, we make contact with the
transverse-momentum-dependent QCD factorization involving the
so-called Sivers functions. We find that the two mechanisms,
quark-gluon correlations on one hand and the Sivers effect on the
other, contain the same physics in this region. This ties the two
together and imposes an important constraint on phenomenological
studies of single transverse spin asymmetries.
\end{abstract}

\maketitle

\newcommand{\be}{\begin{equation}}
\newcommand{\ee}{\end{equation}}
\newcommand{\ben}{\[}
\newcommand{\een}{\]}
\newcommand{\beqn}{\begin{eqnarray}}
\newcommand{\eeqn}{\end{eqnarray}}
\newcommand{\Tr}{{\rm Tr} }

\section{Introduction}

Single-transverse spin asymmetries (SSAs) play an important role
for our understanding of QCD and of nucleon structure. They have a
long history, starting from the 1970s and 1980s when surprisingly
large SSAs were observed in hadronic reactions such as
$p^{\uparrow}p\to \pi X$ at forward angles of the produced pion
\cite{E704-Bunce}. The last few years have seen a renaissance in
the experimental studies of SSAs. The HERMES collaboration at
DESY, SMC and COMPASS at CERN, and the CLAS collaboration at the
Jefferson Laboratory have investigated SSAs in semi-inclusive
hadron production $eN^{\uparrow}\to e \pi X$ in deep-inelastic
scattering \cite{dis}. For proton targets, remarkably large
asymmetries were found. The advent of the first polarized
proton-proton collider, RHIC, has opened new possibilities for
extending the studies of SSAs in hadronic scattering into a regime
where the use of QCD perturbation theory in the analysis of the
data appears to be justified. The STAR, PHENIX and BRAHMS
collaborations have presented data for single-inclusive hadron
production, and large single-spin effects at forward rapidities
were found to persist to RHIC energies \cite{rhic}.

The observed large size of SSAs in hadronic scattering has
presented a challenge for QCD theorists~\cite{review}. Two
mechanisms have been proposed~\cite{Siv90,Efremov,qiu} and been
extensively
applied~\cite{qiu,Kanazawa:2000hz,Ans94,MulTanBoe,siverscompare}
in phenomenological studies. The first relies on the use of
transverse-momentum dependent parton distributions for the
transversely polarized proton. For these distributions, known as
Sivers functions \cite{Siv90}, the parton transverse momentum is
assumed to be correlated with the proton spin vector, so that spin
asymmetries naturally arise from the directional preference
expressed by that correlation. The other mechanism (referred to as
Efremov-Teryaev-Qiu-Sterman (ETQS) mechanism) is formulated in
terms of the collinear factorization approach and twist-three
transverse-spin-dependent quark-gluon correlation functions of the
proton \cite{Efremov,qiu}.

A concept common to both mechanisms is the factorization of the
spin-dependent cross section into functions describing the
distributions of quarks and gluons in the polarized proton, and
partonic hard-scattering cross sections, calculated in QCD
perturbation theory. The question of which mechanism should be
used in the analysis of a single-spin asymmetry is therefore
primarily tied to the factorization theorem that applies for the
single-spin observable under consideration. For example, for the
single-inclusive process $p^{\uparrow}p\to \pi X$, there is only
one hard scale, the transverse momentum $p_T$ of the produced
pion, and the SSA is power-suppressed (``higher-twist'') by
$1/p_T$. In this case, one can prove a {\it collinear}
factorization theorem in terms of the quark-gluon correlation
functions~\cite{qiu}, and the ETQS mechanism applies. On the other
hand, the observables typically investigated in deep-inelastic
lepton scattering (DIS) are characterized by a large scale $Q$
(the virtuality of the DIS photon) and by the much smaller, and also
measured, transverse momentum $q_\perp$ of the produced hadron. In
this two-scale case, single-spin asymmetries may arise at leading
twist, i.e., {\it not} suppressed by $1/Q$. The relevant
factorization theorem is formulated in terms of {\it
transverse-momentum-dependent} (TMD) functions
\cite{ColSop81,ColSopSte85,JiMaYu04,ColMet04}, in particular the
Sivers functions. We note that much progress has been made
recently in understanding the underlying theoretical issues in the
TMD QCD factorization. For example, the gauge-invariance
properties of the necessary TMD parton distributions have been
clarified
\cite{BroHwaSch02,Col02,BelJiYua02,BoeMulPij03} for DIS and Drell-Yan
processes, and an extension to more complicated hadronic processes
has been considered in \cite{Pijl}.

From what we just discussed, the two mechanisms for single-spin
asymmetries might appear to be essentially unrelated. However, one can
make an argument that a consistent theoretical description of the SSA for
a hard process over its full kinematical regime requires both
mechanisms to be present and to contain the same physics in the
region where they both apply. To give a specific example -- which
will in fact be the example treated in this paper -- let us
consider the SSA for the Drell-Yan process when the invariant mass
$Q$ of the pair as well as its transverse momentum $q_\perp$ are
measured. We stress that we consider cross sections differential
just in $Q^2$ and $q_\perp$, but not specifically in the angular
distributions of any of the individual leptons.

At relatively large pair transverse momentum, $q_\perp\sim Q$,
there is only one large scale, and the SSA will be
power-suppressed in that scale. This directs us to use the ETQS
mechanism with its collinear factorization involving the
twist-three quark-gluon correlation functions and corresponding
hard-scattering functions calculated at lowest order from partonic
$3\to 2$ processes. As an aside, when taking the real-photon
($Q^2\to 0$) limit, this calculation will yield the result for the
SSA in direct-photon production~\cite{qiu}.

We can next investigate what happens when we make the ratio
$q_\perp/Q$ small, keeping both scales perturbative, $Q\gg q_\perp
\gg \Lambda_{\rm QCD}$. We refer to $q_\perp$ in this regime as
``moderate'' transverse momentum. The ETQS mechanism will still
apply here (even though the hard-scattering functions will develop
large logarithms of the ratio $q_\perp/Q$ that will eventually
need to be resummed to all orders in the strong coupling). At the
same time, however, the factorization in terms of TMD
distributions applies now~\cite{ColSop81,ColSopSte85,JiMaYu04},
which involves the Sivers functions. If both mechanisms are
internally consistent, they must describe the same physics in this
region.

In a recent publication \cite{JiQiuVogYua06}, we have demonstrated
that the two mechanisms indeed provide the same description of the
single-spin asymmetry for the Drell-Yan process in the regime
$\Lambda_{\rm QCD}\ll q_\perp \ll Q$, and that there is a direct
correspondence between the Sivers functions and the twist-three
quark-gluon correlation functions, derived also earlier
in~\cite{BoeMulPij03}. The key observation is that, at moderate
transverse momentum, the Sivers function may be calculated
perturbatively, using the twist-three quark-gluon correlation
functions. In other words, the ETQS mechanism generates a
non-vanishing Sivers function in this kinematic regime.
Although there have been earlier efforts to link the two mechanisms
\cite{BoeMulPij03,MaWa03,bacchetta}, a clear connection,
at the level of physical observables, has been established only
in~\cite{JiQiuVogYua06}. In the present paper, we present details
of the derivation of our results in~\cite{JiQiuVogYua06}.

Our results may in some sense be viewed as establishing a
unification of the two mechanisms. At large $q_\perp$, the ETQS
mechanism applies. At moderate transverse momentum, a smooth
transition from the ETQS mechanism to the one based on TMD
factorization occurs, with the two approaches containing the same
physics. At yet lower $q_\perp$ ($\sim \Lambda_{\rm QCD}$),
the TMD factorization  still applies, containing in a natural way the
transition from perturbative to non-perturbative physics. We
believe that this unified picture should prove to be the best
approach to phenomenological studies of SSAs.

The remainder of the paper is organized as follows. In Sec. II, we
introduce the twist-three quark-gluon correlation function and
explain its physical significance. In Sec. III, we calculate the
single-spin differential cross section for Drell-Yan production
using the ETQS mechanism, and we study its limit $q_\perp\ll Q$ in
Sec. IV. A comparison with the TMD factorization approach is made
in Sec. V, and we conclude the paper in Sec. VI.

\section{Transverse-Spin-Dependent Quark-Gluon Correlation}

Consider a transversely polarized proton traveling at nearly the
speed of light. Its internal color electric and magnetic fields
then have preferred orientations in the plane transverse to the
proton's direction of motion. By parity invariance, the color
electric field must be orthogonal to the spin of the proton. If
averaged over the proton wave function, the field vanishes because
the proton is color-neutral (and also because of time-reversal
symmetry). However, if one multiplies the color field with the
quark color current, the average may be non-zero. This average
defines a quark-gluon correlation function that characterizes a
property of a polarized proton. In some sense, the correlation
describes how strongly polarized the color electric field is in a
spinning proton.

In this paper, we are mostly interested in the so-called
light-cone correlations. For these, quark and/or gluon fields are
separated along the light-cone direction $\xi^-$ (if $\xi^\mu$
denotes a space-time coordinate, the light-cone variables are
defined as $\xi^\pm = (\xi^0\pm\xi^3)/\sqrt{2}, \xi_\perp =
(\xi^1, \xi^2)$). It is the light-cone correlations that
characterize the structure of the proton in high-energy
scattering. One finds the following expression for the
spin-dependent quark-gluon correlation described above~\cite{qiu}:
\begin{eqnarray}
\Phi_F^\alpha(k_{q1}^+,k_{q2}^+)^a_{ij}
&=&\int\frac{d\zeta^-d\eta^-}
{(2\pi)^2}e^{ik_q^+\eta^-}e^{ik_g^+\zeta^-} \left\langle
PS|\overline\psi_{\sigma j}(0){\cal
L}(0,\zeta^-)gF_a^{+\alpha}(\zeta^-) {\cal
L}(\zeta^-,\eta^-)\psi_{\rho i}(\eta^-)|PS\right\rangle
\nonumber\\
&=& \frac{1}{2\pi}\frac{1}{2}(\gamma_\mu)_{\rho\sigma} P^\mu\:
\epsilon_\perp^{\beta\alpha}S_{\perp\beta}\:
T_F(x_1,x_2)\frac{T^a_{ij}}{N_C C_F} + \ldots \ , \label{Phidef}
\end{eqnarray}
visualized by the diagram in Fig.~1.
\begin{figure}
\begin{center}
\includegraphics[height=3.0cm]{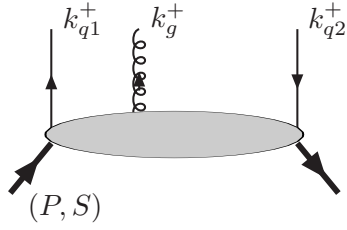}
\end{center}
\caption{\it A vertex diagram representing twist-three quark-gluon
correlation function in a proton. When the proton is polarized,
the gluon field has a definite polarization.}
\end{figure}
Here, $|PS\rangle$ is the proton state with four-momentum $P^\mu$
and transverse-polarization vector $S^\mu= (0,0,\vec{S}_\perp)$,
$\bar{\psi}$, $\psi$ are the quark fields with color indices $i,j$
and Dirac indices $\rho,\sigma$, and $F^{+\alpha}_a$ is the gluon
field strength tensor with octet color index $a$.
$x_1=k_{q1}^+/P^+$, $x_g=k_g^+/P^+$, and
$x_2=k_{q2}^+/P^+=x_1+x_g$ are the fractions of the proton's
light-cone momentum carried by the quarks and the gluon, where
their momenta as shown in Fig.~1. Furthermore, ${\cal L}$ is the
light-cone gauge link,
\begin{equation}
   {\cal L}(\xi_2^-, \xi_1^-) =
   P\exp\left(-ig\int^{\xi_2^-}_{\xi_1^-} d\xi^-
   A^+(\xi^-)\right) \ ,
\end{equation}
which makes the correlation operator gauge-invariant. Note that
our sign convention for the strong coupling constant $g$ follows
from $D^\mu \equiv \partial^\mu +igA^\mu$ for the covariant
derivative. Also note that we have included a factor $g$ in the
matrix element in Eq.~(\ref{Phidef}), as compared to the
definition in~\cite{qiu}. In the second line of (\ref{Phidef}), we
have expanded the matrix in Dirac space, keeping only the leading,
twist-three, term that contributes to the SSA in the Drell-Yan
process, and neglecting contributions of yet higher twist. Here,
$N_C=3$ is the number of colors, $C_F=(N_C^2-1)/2N_C=4/3$, and
$T^a_{ij}$ denote the generators of the SU(3) color gauge group.
$\epsilon^{\alpha\beta}_\perp$ is the 2-dimensional Levi-Civita
tensor with $\epsilon^{12}_\perp=1$.

Inverting Eq. (1), we can express the correlation function
$T_F(x_1, x_2)$ as follows:
\begin{equation}
T_F(x_1,x_2)=
\int\frac{d\zeta^-d\eta^-}{4\pi}e^{ix_1P^+\eta^-}e^{i(x_2-x_1)P^+\zeta^-}
\epsilon_{\perp}^{\beta\alpha}S_{\perp\beta}\left\langle
PS|\overline\psi(0)\gamma^+g{F^{+}}_{\alpha}(\zeta^-)\psi(\eta^-)|PS\right\rangle
\ , \label{Tdef}
\end{equation}
where the sums over color and spin indices are now implicit and,
for simplicity, we have omitted the gauge link which vanishes if
one chooses to work in the light-cone gauge, $A^+=0$. Note that we
have suppressed a renormalization scale dependence of $T_F$.
Equation~(\ref{Tdef}) provides the normalization for our
calculations in the following sections. Because of parity and
time-reversal invariance, the correlation function has the
symmetry property~\cite{qiu}
\begin{equation}
T_F(x_1,x_2)=T_F(x_2,x_1) \label{symm}\ ,
\end{equation}
which will be used later on to simplify our results.

\section{Single-Spin Drell-Yan Cross Section}

We shall now calculate the single-transverse-spin dependent cross
section for the Drell-Yan process, using the ETQS mechanism. We
will first perform the calculation of the cross section at parton
level. Convolution with the twist-three correlation function
introduced in the last section and with the usual Feynman parton
distributions for the unpolarized proton will then give the
physical hadronic cross section. We shall limit ourselves to a
situation where the dominant contributions come from the
twist-three correlation function for a {\it valence quark}, and
from an anti-quark or gluon of the unpolarized proton. In this
way, we do not need to worry, for example, about other types of
twist-three correlation functions, such as purely gluonic ones
\cite{Ji:1992eu}. In an actual experiment, such a situation may be
realized when the lepton pair is produced in the forward direction
of the polarized beam (at large rapidity), so that large momentum
fractions in the polarized proton are probed.

As we discussed in the Introduction, we are primarily interested
here in the Drell-Yan pair produced at large transverse momentum
$q_\perp\sim Q$. For this to happen, a quark or gluon jet must be
produced in the hard partonic process against which the Drell-Yan
pair recoils. In the unpolarized case, to lowest order in
perturbation theory, there are two partonic subprocesses of this
kind: the Compton-type quark-gluon scattering $q+g\rightarrow
\gamma^*+q$, and quark-antiquark annihilation $q+\bar q\rightarrow
\gamma^*+g$. One therefore arrives at the following lowest-order
(LO) expression of the unpolarized Drell-Yan cross section for
producing a virtual photon of invariant mass $Q$, rapidity $y$,
and transverse momentum $q_\perp$:
\begin{eqnarray}
\frac{d^4\sigma}{dQ^2dyd^2q_\perp}&=&\sigma_0 \frac{\alpha_s
}{4\pi^2}\int \frac{dx}{x}\frac{dx'}{x'} \sum_q e_q^2\left[
\hat{\sigma}_{q\bar{q}}(\hat{s},\hat{t},\hat{u}) q(x)\bar{q}(x')
+\hat{\sigma}_{qg}(\hat{s},\hat{t},\hat{u})
q(x)g(x')\right]\nonumber \\
&&\,\times\delta(\hat s+\hat t+\hat u-Q^2) \label{sigunp} \ ,
\end{eqnarray}
where $\alpha_s$ is the strong coupling constant, and $\sigma_0 =
4\pi\alpha_{\rm em}^2/3N_C sQ^2$ with the electromagnetic coupling
$\alpha_{\rm em}$ and the hadronic center-of-mass energy squared
$s=(P+P')^2$. Furthermore, $x$ and $x'$ are the partons' momentum
fractions, and the partonic Mandelstam variables are defined as
$\hat s=(xP+x'P')^2$, $\hat t=(xP-q_{\gamma^*})^2$, $\hat
u=(x'P'-q_{\gamma^*})^2$ with the virtual photon momentum
$q_{\gamma^*}$. Note that we have omitted the scale dependence of
the parton distributions, as we will do throughout this paper.
\begin{figure}[t]
\begin{center}
\includegraphics[height=5.0cm]{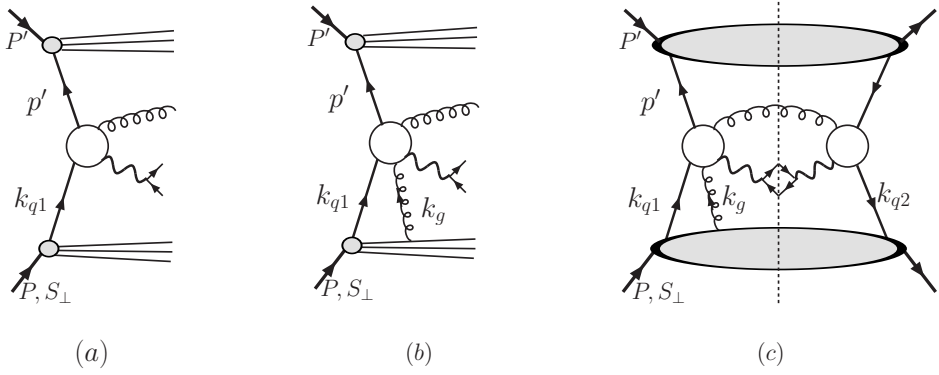}
\end{center}
\caption{{\it Drell-Yan scattering amplitudes via (a)
$q+\bar{q}\rightarrow \gamma^* + g$ and (b) $q+\bar{q} + g
\rightarrow \gamma^* + g$; (c) A typical diagram, from the
interference of the amplitudes in (a) and (b), that gives a
contribution to the SSA.}}
\end{figure}

In order to have a non-vanishing single-transverse-spin asymmetry
in a hadronic process, an interference of two amplitudes with
different strong-interaction phases is required. When the
underlying scattering mechanism is hard, as is the case for our
Drell-Yan observable at large $Q$ and $q_\perp$, the difference in
strong interaction phase may arise from the interference between a
real part of the scattering amplitude in Fig.~2(a) and an
imaginary part of the partonic scattering amplitude with an extra
gluon in Fig.~2(b). The interference of these two amplitudes corresponds
to the typical diagram contributing to the SSA shown in Fig.~2(c).  The
additional gluon of momentum $k_g$ from the polarized proton,
which leads to the twist-three quark-gluon correlation function as
shown in Fig.~1, can attach to any propagator of the hard part
represented by a light circle in Fig.~2(b) or (c). The imaginary
part of the amplitude with an extra gluon is provided by the pole
of the parton propagator associated with the integration of the
gluon momentum fraction $x_g=k_g^+/P^+$~\cite{qiu}. In the
previously considered cases of the SSAs in direct-photon or
inclusive-hadron production, such poles only occur when the
additional polarized gluon has a vanishing momentum, $x_g=0$
($x_1=x_2$), while it is attached to an external on-shell
parton~\cite{qiu}. These poles were referred to as ``soft poles''
\cite{luo,guo}. For example, if the polarized gluon attaches to
the incoming unpolarized anti-quark, the on-shell propagation of
the anti-quark line will generate such a soft and un-pinched
gluonic pole. However, for the Drell-Yan process, there are two
observed hard scales, $Q$ and $q_\perp$, because of the outgoing
photon being off-shell, and there could be additional imaginary
contributions associated with poles of propagators in the partonic
scattering amplitude for which $x_g\ne 0$ ($x_1\ne x_2$). 
Such poles were called ``hard poles'' \cite{luo,guo}. In the
calculations presented below, we shall consistently include both
contributions, soft-pole and hard-pole ones. Compared to the
unpolarized case in Eq.~(\ref{sigunp}), we now find the following
structure for the LO single-spin cross section:
\begin{eqnarray}
\frac{d^4\Delta\sigma(S_\perp)}{dQ^2dyd^2q_\perp}&=&
\sigma_0\epsilon^{\alpha\beta}{S}_{\perp\alpha}
q_{\perp\beta}\frac{\alpha_s}{2\pi^2}\int\frac{dx}{x}\frac{dx'}{x'}
\sum_q e_q^2\left[ \left(H_q^s+H_q^h\right) \bar{q}(x')
+\left(H_g^s+H_g^h
\right)g(x')\right]\nonumber \\
&&\times \,\delta(\hat s+\hat t+\hat u-Q^2) \ , \label{quarkform}
\end{eqnarray}
where $H_{q,g}^s$ and $H_{q,g}^h$ are the soft- and hard-pole
contributions for scattering off an unpolarized anti-quark or
gluon, respectively. Each of these will be a function of the
partonic Mandelstam variables, of $x$ and $x'$, and of the
twist-three quark-gluon correlation function $T_F(x_1,x_2)$. Note
that $x$ is the same as in the unpolarized case, determined in
terms of the external hadronic variables through the
delta-function in~(\ref{quarkform}). As will be described in
detail below, $x_1$ and $x_2$ depend on $x$ and are fixed by the
on-shell conditions set variously by the soft and hard poles, and
by that for the unobserved final-state parton in the hard process.

In the following subsections, we will first discuss the generic
features of the soft and hard poles, and then present their
respective contributions to the single-transverse polarized cross
sections for the quark-antiquark and quark-gluon scattering
channels.

\subsection{Generic Structure of Soft and Hard Poles}

\begin{figure}[t]
\begin{center}
\includegraphics[height=4.0cm]{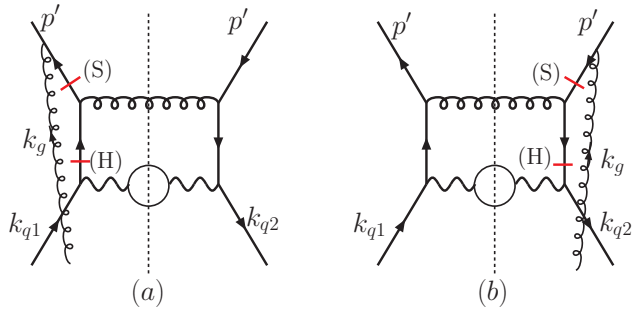}
\end{center}
\vskip -0.7cm \caption{\it A typical Feynman diagram (a) and its
complex conjugate (b) contributing to the single-spin asymmetry
for the Drell-Yan process through the quark-antiquark scattering
channel, containing both soft-pole ``(S)'' and hard-pole ``(H)''
contributions. The bars indicate the places where a pole arises
and generates a strong interaction phase.}
\end{figure}

We will take the diagrams shown in Fig.~3 as examples of how the
soft and hard poles arise. Let $p'$ denote the momentum of the
incident anti-quark, $k_{q1}$ that of the initial quark to the left
of the cut, $k_{q2}$ that on the right, and $k_g=k_{q2}-k_{q1}$
the momentum of the polarized gluon attaching to the hard part.
This attachment may happen on the left side of the cut, as shown
in Fig~3(a), or on the right side as in 3(b). In order to analyze
the pole structure in the scattering amplitudes, it is convenient
to consider only the dominant components of the relevant momenta.
For example, $p'$ may be chosen to only have a light-cone
``minus'' component, while $k_{q1}$ and $k_{q2}$ are then
dominated by their plus components. When the polarized gluon
attaches to the left side of the cut in Fig.~3(a), we find that
the on-shell condition for the gluon radiated into the final state
fixes $k_{q2}^+$ (or, equivalently, $x_2$) to be $x_2\approx x$,
while $k_{q1}^+$ ($x_1$) is determined by the (soft- or hard-)
pole condition. On the other hand, the momentum flow in Fig.~3(b)
is opposite: $x_1$ is determined by the on-shell condition for the
outgoing gluon, and $x_2$ by the pole of a propagator.

In Fig.~3, we have denoted the poles by bars, and introduced the
labels $(S)$ and $(H)$ for the soft and hard poles, respectively.
A soft pole arises in Fig.~3(a) on the anti-quark line carrying
momentum $p'+k_g$, because we have
\begin{equation}
\frac{1}{(p'+k_g)^2+i\epsilon}=\frac{1}{2p^{'-}P^+}\frac{1}{x_g+i\epsilon}
\ , \label{3a}
\end{equation}
which provides a phase proportional to $\delta
(x_g)=\delta(x_1-x_2)$. Therefore, the quark-gluon correlation
function $T_F(x_1,x_2)$ will be probed when both of its arguments
are identical. After combining with the on-shell condition for the
unobserved final-state gluon, one finds in fact that the
contribution enters with $T_F(x,x)$~\footnote{ As we shall discuss
later, there are also contributions with the derivative
$dT_F(x,x)/dx$.}. Similarly, the soft pole in Fig.~3(b) will arise
through
\begin{equation}
\frac{1}{(p'-k_g)^2-i\epsilon}=\frac{1}{2p^{'-}P^+}\frac{1}{-x_g-i\epsilon}
\label{3b} \ ,
\end{equation}
so that it enters with opposite sign compared to that of
Fig.~3(a), but with otherwise the same quark-gluon correlation
function. Therefore, the contributions by the two poles in
Eqs.~(\ref{3a}),(\ref{3b}) may be combined.

Next, we turn to the hard pole in Fig.~3(a). It occurs when the
quark propagator carrying momentum $k_{q1}-q_{\gamma^*}$ goes on
mass-shell. The propagator reads
\begin{equation}
\frac{1}{(k_{q1}-q_{\gamma^*})^2+i\epsilon}\approx
\frac{1}{-2k_{q1}^+q_{\gamma^*}^-+Q^2+i\epsilon}\ ,
\end{equation}
where we have neglected the transverse momentum $k_{q1\perp}$. It
is easy to see that this propagator has a pole at $k_{q1}^+=x
P^+{Q^2}/{(Q^2-\hat t)}$, corresponding to $x_g=\bar x_g$, where
\begin{equation}
\bar x_g\equiv -x \hat t/(Q^2-\hat t) \ .
\end{equation}
Because $\hat{t}=(xP-q_{\gamma^*})^2<0$, this pole is not
forbidden kinematically. Furthermore, $x_g>0$, which is the reason
why we refer to the pole as ``hard''. After taking the pole, we
find $x_1=x-\bar x_g$. Using the on-shell condition for the
outgoing gluon, the associated twist-three correlation eventually
is $T_F(x-\bar x_g,x)$, different from that for the soft-pole
contributions. Likewise, the hard pole in Fig.~3(b) arises from
the propagator
\begin{equation}
\frac{1}{(k_{q2}-q_{\gamma^*})^2-i\epsilon}\approx
\frac{1}{-2k_{q2}^+q_{\gamma^*}^-+Q^2-i\epsilon}\ ,
\end{equation}
and is at $k_{q2}^+=x P^+{Q^2}/{(Q^2-\hat t)}$, corresponding to
$x_g=-\bar x_g=x \hat t/(Q^2-\hat t)<0$. The associated
twist-three quark-gluon correlation function becomes $T_F(x,x-\bar
x_g)$. Because of the symmetry property~(\ref{symm}) of the
function, we have $T_F(x,x-\bar x_g)=T_F(x-\bar x_g,x)$, and we
may again combine the contributions from the two sides of the cut
in the diagram.

From kinematics, we find that hard poles only arise from
$t$-channel propagators, and only on the side of the diagram that
contains the additional gluon from the polarized proton. For each
diagram, we have to include all contributions by the soft and hard
poles, in order to obtain the complete result. Although we will
calculate their contributions separately, the soft and hard poles
do overlap in some kinematical regions, for example, when the
transverse momentum $q_\perp$ is much smaller than $Q$. We will
discuss this more in Sec. IV.

\subsection{Soft-Pole Contributions}

In this subsection, we calculate the soft gluonic pole
contributions to the single-transversely-polarized Drell-Yan cross
section in the LO quark-antiquark and quark-gluon scattering
channels. The calculation is rather similar to that performed for
direct-photon production in Ref.~\cite{qiu}, except for the fact
that the photon is now virtual.

\subsubsection{Soft Poles in Quark-Antiquark Scattering}

We start by recalling the unpolarized hard-scattering function for
the subprocess $q+\bar q\rightarrow \gamma^*+g$, appearing in
Eq.~(\ref{sigunp}):
\begin{eqnarray}
\hat{\sigma}_{q\bar{q}} (\hat{s},\hat{t},\hat{u})= 2 C_F \left(
\frac{\hat u}{\hat t}+\frac{\hat t}{\hat u}+\frac{2Q^2\hat s}{\hat
u\hat t}\right) \ ,\label{sigqqunp}
\end{eqnarray}
where $C_F$ is the color-factor and the partonic Mandelstam
variables are as defined after Eq.~(\ref{sigunp}) at the beginning
of this section.

\begin{figure}[t]
\begin{center}
\includegraphics[height=8.0cm]{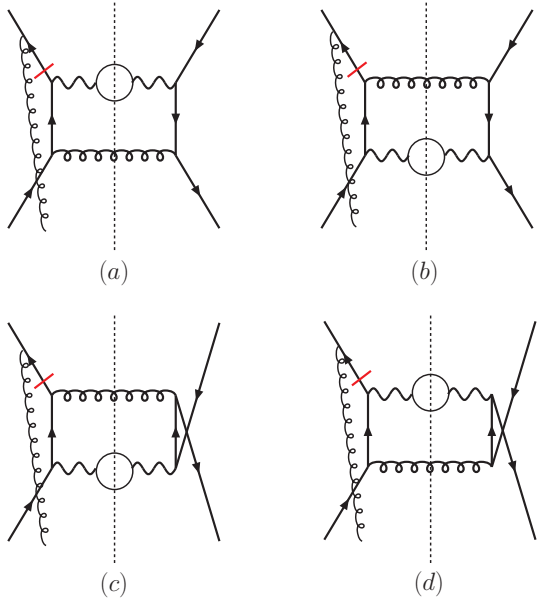}
\end{center}
\vskip -0.7cm \caption{\it Feynman diagrams for the soft-pole
contributions to the single-transverse-spin asymmetry through
the quark-antiquark scattering channel. The bars indicate the
propagators yielding poles.}
\end{figure}

For the single-spin cross section in this channel, there are eight
partonic diagrams possessing soft poles. Four of them are shown in
Fig.~4, and the other four are obtained by attaching the polarized
gluon in the same way on the right side of the cut. Other
attachments of the polarized gluon either do not give a soft pole,
or simply cancel after summing over cuts.

It is useful to recapitulate some of the key steps in calculating
the diagrams in Fig. 4 \cite{qiu}. We perform our calculations
using a covariant gauge. The polarized gluon is associated with a
gauge potential $A^\mu$, and one of the leading contributions
comes from its component $A^+$. The gluon's momentum is dominated
by $x_gP+k_{g\perp}$, where $x_g$ is the longitudinal momentum
fraction with respect to the polarized proton. The transverse
momentum $k_{g\perp}$ flows through the perturbative diagram and
returns to the polarized proton through the quark lines. The
contribution to the single-transverse-spin asymmetry arises from
terms linear in $k_{g\perp}$ which, when combined with $A^+$,
yield $\partial^\perp A^+$, a part of the gauge field strength
tensor $F^{\perp +}$. In order to compute this contribution, we
expand the partonic scattering amplitudes in terms of $k_{g\perp}$
up to the linear term. One important contribution to the
$k_{g\perp}$ expansion comes from the on-shell condition for the
outgoing unobserved quark, whose momentum depends on $k_{g\perp}$.
As was shown in~\cite{qiu}, this leads to a term involving the
{\it derivative} of the correlation function $T_F$. Apart from
this contribution, the $k_{g\perp}$ expansion of the propagators
and the Dirac traces yields a term proportional to the correlation
function itself. This is the so-called non-derivative term.

The sum of the soft-pole contributions by the diagrams in Fig.~4
and their ``mirror images'' gives the function $H_q^s$ in
Eq.~(\ref{quarkform}). We find:
\begin{eqnarray}
H_q^s &=& \left[x\frac{\partial}{\partial x}T_F(x,x)\right]
\frac{D^s_{q\bar q}}{-\hat u}+T_F(x,x) \frac{N^s_{q\bar q}}{-\hat
u}\ , \label{hq}
\end{eqnarray}
where the hard coefficients $D^s_{q\bar q}$ and $N^s_{q\bar q}$
are
\begin{eqnarray} D^s_{q\bar q}&=&
\frac{1}{2 (N_C^2-1)}\,\hat{\sigma}_{q\bar{q}}
(\hat{s},\hat{t},\hat{u})\ ,
\nonumber\\
N^s_{q\bar q}&=&\frac{1}{2N_C}\frac{1}{\hat t^2\hat u}\left[
Q^2(\hat u^2-\hat t^2)+2Q^2\hat s(Q^2-2\hat t)-(\hat u^2+\hat
t^2)\hat t\right]\nonumber\ ,
\end{eqnarray}
where $\hat{\sigma}_{q\bar q}$ in the first line is the
unpolarized hard-scattering function given in
Eq.~(\ref{sigqqunp}), implying that the hard-scattering function
for the derivative term differs from the unpolarized one only by
the color factor~\cite{Ratcliffe}. In the real-photon limit,
$Q^2\to 0$, we obtain the annihilation contribution to the
direct-photon single-spin cross section.

\subsubsection{Soft Poles in Quark-Gluon Scattering}

Again we start by giving the hard-scattering function for the
unpolarized case:
\begin{eqnarray}
\hat{\sigma}_{qg}(\hat{s},\hat{t},\hat{u}) =  2 T_R \left(
\frac{\hat s}{-\hat t}+ \frac{-\hat t}{\hat s}-\frac{2Q^2\hat
u}{\hat s\hat t} \right) \, , \label{sigqgunp}
\end{eqnarray}
where $T_R=1/2$.
\begin{figure}[t]
\begin{center}
\includegraphics[height=8.0cm]{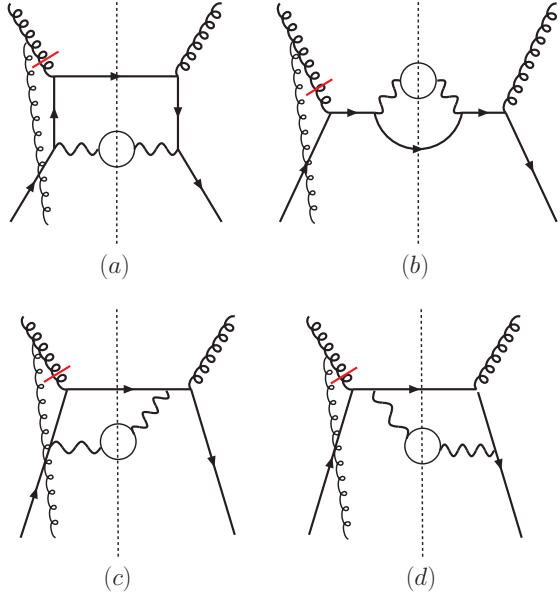}
\end{center}
\vskip -0.7cm \caption{\it Feynman diagrams for soft-pole
contributions to the single-transverse-spin asymmetry through
quark-gluon scattering.}
\end{figure}

Feynman diagrams for the soft-pole contributions to the
single-transverse-spin asymmetry are shown in Fig.~5, where again the
diagrams for the attachments of the polarized gluon to the
incident gluon on the right side of cut have been omitted.
Following the same calculational procedure as outlined earlier,
the soft-pole contributions for the quark-gluon channel are:
\begin{eqnarray}
H_g^s &=& \left[x\frac{\partial}{\partial x}T_F(x,x)\right]
    \frac{D^s_{qg}}{-\hat u}+T_F(x,x) \frac{N^s_{qg}}{-\hat u}\ ,
\end{eqnarray}
where the first term is the derivative term and the second the
non-derivative one, with
\begin{eqnarray}
D^s_{qg}&=&-\frac{N_C^2}{2(N_C^2-1)}\,\hat{\sigma}_{qg}(\hat{s},\hat{t},\hat{u})\
,
\nonumber\\
N^s_{qg}&=&\frac{N_C^2}{2(N_C^2-1)}\frac{1}{\hat t^2\hat s}\left[
Q^2(\hat s^2-\hat t^2)+2Q^2\hat u(Q^2-2\hat t) -(\hat s^2+\hat
t^2)\hat t\right]\nonumber\ .
\end{eqnarray}
We note that apart from the color factor, the hard coefficients
$D^s_{qg}$ and $N^s_{qg}$ for the quark-gluon scattering channel can be
obtained from those for the quark-antiquark channel by
``crossing'' $\hat s\leftrightarrow \hat u$. In the limit $Q^2=0$,
we obtain the direct-photon cross section, which was calculated
previously in~\cite{qiu}. We agree with the derivative term found
there, but find a difference for the non-derivative one.

\subsection{Hard-Pole Contributions}

In this subsection, we will calculate the hard-pole contributions
to the single-spin cross section. The calculational procedure is
similar to that for the soft-pole contributions in the previous
subsection. We will again discuss the quark-antiquark and
quark-gluon channels separately.

\subsubsection{Hard Poles in Quark-Antiquark Scattering}

\begin{figure}[t]
\begin{center}
\includegraphics[height=8.0cm]{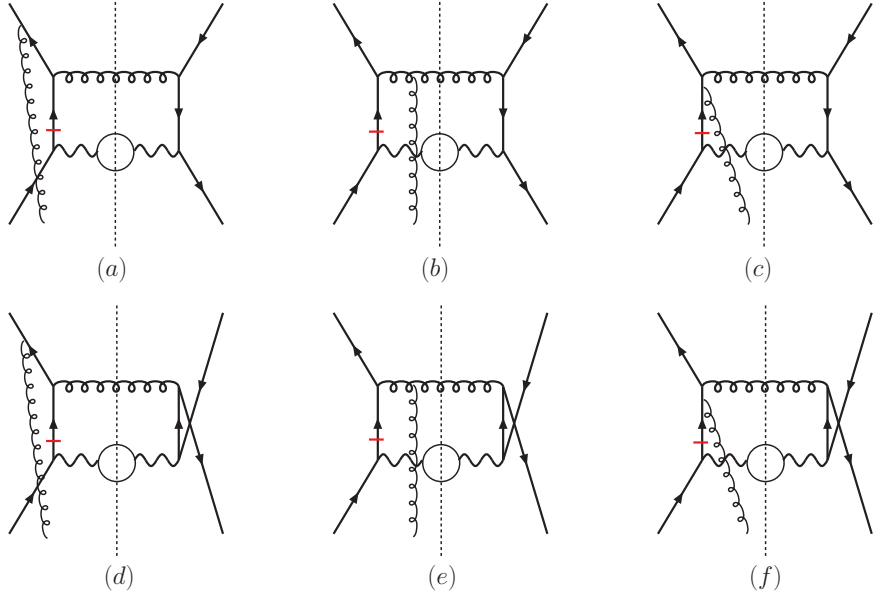}
\end{center}
\vskip -0.7cm \caption{\it Hard-pole contributions in the
quark-antiquark scattering channel. The bar indicates the
propagator that produces the pole.}
\end{figure}

In Fig.~6 we show the hard-pole diagrams for the quark-antiquark
scattering channel, again omitting the ``mirror'' diagrams. As
discussed in subsection III.A, only $t$-channel propagators give a
hard pole. In the quark-antiquark scattering channel, there are
only two basic diagrams containing a $t$-channel propagator, shown
in Figs.~6(a) and (d). However, because of gauge invariance, we
need to allow all possible gluon attachments in these diagrams. We
thus have attachments to the initial antiquark line, to the
outgoing gluon line and to the $t$-channel quark propagator, as
shown in the figure. Therefore, including the diagrams with gluon
attachments on the right side of the cut, there will be a total of
twelve diagrams containing hard poles in the quark-antiquark
scattering channel. We may group their contributions into two
color structures: ${\rm Tr}(T^aT^bT^aT^b)$ for diagrams (a) and
(d), and ${\rm Tr}(T^aT^aT^bT^b)$ for diagrams (c) and (f). The
first color factor is the same as for the soft gluonic pole
diagrams. The color traces for diagrams (b) and (e) can be
decomposed into these two color structures, and combined with
(a,d) and (c,f), using the fact that $if_{abc}T^c=T^aT^b-T^bT^a$.

The expansion in $k_{g\perp}$ for the hard-pole contributions is
somewhat more complicated than for the soft-pole ones, because the
position of the hard pole itself depends on $k_{g\perp}$. As an
example we consider the diagram in Fig.~6(a), for which a hard
pole arises from the propagator
\begin{equation}
 \frac{1}{(k_{q1}-q_{\gamma^*})^2+i\epsilon}
=\frac{1}{-2k_{q1}^+q_{\gamma^*}^-
          +Q^2-2k_{q1\perp}\cdot q_{\gamma^*\perp}+i\epsilon}\ ,
\end{equation}
where we have kept the full dependence on the transverse momentum.
In the $k_{g\perp}$-expansion, the $k_{q1\perp}$ term in the above
denominator will make a contribution to the term linear in
$k_{g\perp}$, because $k_{g\perp}=k_{q2\perp}-k_{q1\perp}$. This
will lead to a derivative term (of $T_F$), similar to the
double-pole contributions found in~\cite{qiu} for the SSA in
single-inclusive hadron production. In the calculation of this
derivative term, we may set $k_{g\perp}=0$ everywhere else in the
scattering amplitude. Because all the propagators with hard poles
in Fig.~6 are $t$-channels and hence have the same momentum,
namely $k_{q1}-q_{\gamma^*}$, the expansion will be the same for
all diagrams, and we may add their contributions. After summing
over all the diagrams, we find that the total derivative
contribution vanishes. More precisely, the derivative
contributions from the hard poles in diagrams (a), (b), (d) and
(e) (which all contain the color structure ${\rm Tr}
(T^aT^bT^aT^b)$) cancel each other, and the same happens for the
contributions by diagrams (b), (c), (e) and (f) (which enter with
${\rm Tr}(T^aT^aT^bT^b)$).

The expansion of the delta function associated with the on-shell
condition for the outgoing unobserved gluon
($(k_{q2}+p'-q_{\gamma^*})^2=0$) also contributes to the
derivative term for each individual diagram. However, after
summing over all diagrams, again the net derivative contribution
vanishes. In summary, we do not have any derivative contributions
originating from the hard poles.

For the non-derivative contributions, we need to do an expansion
of the numerator (for example, the Dirac trace) and of the other
propagators in the squared amplitude. Apart from that, the
derivative terms just discussed will make contributions to the
non-derivative pieces. Our way of dealing with the non-derivative
contributions is to first evaluate the scattering amplitudes with
their full dependence on $k_{g\perp}$, and then to use the
on-shell condition for the outgoing unobserved gluon and the
hard-pole condition, in order to fix the light-cone momentum
fractions of the quarks including their dependence on
$k_{g\perp}$. For example, the on-shell condition for
$k_{q2}+p'-k_{\gamma^*}$ in Fig.~6(a) determines the value for
$k_{q2}^+$:
\begin{equation}
k_{q2}^+=xP^+\left(1-\frac{2k_{q2\perp}.q_\perp}{\hat u}\right)\ .
\end{equation}
Similarly, the hard-pole condition $(k_{q1}-k_{\gamma^*})^2=0$
fixes the value for $k_{q1}^+$ as
\begin{equation}
k_{q1}^+=xP^+\frac{Q^2}{Q^2-\hat
t}\left(1-\frac{2k_{q1\perp}.q_\perp}{Q^2}\right) \ .
\end{equation}
From the above equations, we see that both $k_{q1}^+$ and
$k_{q2}^+$ depend on $k_{g\perp}$, so that their expansion will
contribute to a non-derivative term.

Summing over all the contributions by the diagrams in Fig.~6 and
their ``mirror images'', we obtain the final result for the
non-derivative hard-pole contribution for the quark-antiquark
channel:
\begin{eqnarray}
H_q^h=T_F(x-\bar x_g,x) \times \frac{(Q^2-\hat t)^3+Q^2\hat
s^2}{\hat t^2\hat u^2} \left[\frac{1}{2N_C}+C_F\frac{\hat s}{\hat
s+\hat u}\right] \ ,
\end{eqnarray}
where as before $\bar x_g=-x\hat t/(Q^2-\hat t)$ is different from
zero, reflecting the hard-pole nature of the contribution. The two
color factors in the above equation are associated with the two
independent color structures discussed earlier, the first one
coming from ${\rm Tr}(T^aT^bT^aT^b)$ and the second one from ${\rm
Tr}(T^aT^aT^bT^b)$.

\subsubsection{Hard Poles in the Quark-Gluon Channel}
\begin{figure}[t]
\begin{center}
\includegraphics[height=8.0cm]{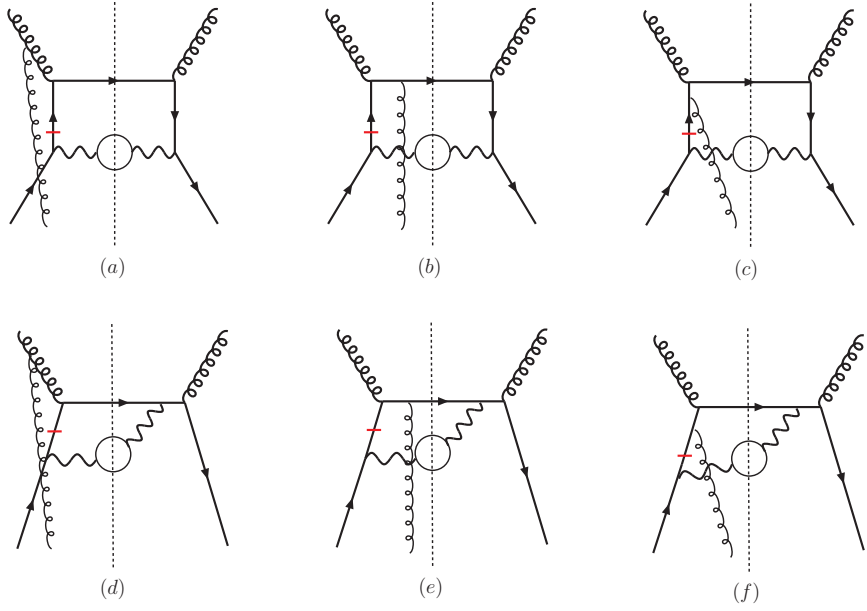}
\end{center}
\vskip -0.7cm \caption{\it Hard pole contributions in the
quark-gluon scattering channel. The bar indicates the propagator
that produces the pole.}
\end{figure}

By a similar procedure, we calculate the hard-pole contributions
for the quark-gluon scattering channel. In Fig.~7, we show the
relevant diagrams. Following the same arguments as given for the
quark-antiquark scattering channel, only $t$-channel diagrams
contribute to the hard poles, and their total contribution is
\begin{eqnarray}
H_g^h=T_F(x-\bar x_g,x)\times \frac{(Q^2-\hat t)^3+Q^2\hat
u^2}{-\hat u\hat t^2\hat
s}\left[\frac{-N_C^2}{2(N_C^2-1)}+T_R\frac{\hat s}{\hat s+\hat
u}\right] \ .
\end{eqnarray}
Again, there are no derivative terms, and the hard coefficient for
the non-derivative term is related to that for quark-antiquark
scattering given above by $\hat s\leftrightarrow \hat u$ crossing.
As for the $q\bar q$ case, there are two contributing color
structures, one associated with $if_{abc}{\rm Tr}(T^aT^bT^c)$ (for
diagrams Fig.~7(a), (b), (d), and (e)), and one with ${\rm
Tr}(T^aT^aT^bT^b)$ (for diagrams Fig.~7(c), (b), (f), and (e)).

\section{Low Transverse Momentum Limit}

Having calculated the full LO single-spin cross section for the
Drell-Yan process at large $q_\perp$ within the ETQS approach, we
are now in the position to investigate the limit $\Lambda_{\rm
QCD}\ll q_\perp \ll Q$, trying to make contact with the approach
based on transverse-momentum dependent factorization. To this end,
we expand the formulas given in the preceding section for small
$q_\perp/Q$, extracting their leading contributions.

The partonic Mandelstam variables may be written as
\begin{eqnarray}
\hat s&=&\frac{q_\perp^2}{(1-\xi_1)(1-\xi_2)}\ , \\
\hat t&=&-\frac{q_\perp^2}{1-\xi_2}\ , \\
\hat u&=&-\frac{q_\perp^2}{1-\xi_1}\ ,
\end{eqnarray}
where $\xi_1= z_1/x$ and $\xi_2=z_2/x'$, with $z_1=Q/\sqrt{s} e^y$
and $ z_2=Q/\sqrt{s} e^{-y}$. The phase-space delta-function for
the on-shell condition for the outgoing parton takes the form~\cite{meng}
\begin{eqnarray}
\delta(\hat s+\hat t+\hat u-Q^2)&=&\delta(\hat
s(1-\xi_1)(1-\xi_2)-q_\perp^2)\nonumber\\
&=&\frac{1}{\hat s}\left[\frac{\delta(\xi_2-1)}{(1-\xi_1)_+}+
\frac{\delta(\xi_1-1)}{(1-\xi_2)_+}+\delta(\xi_1-1)\delta(\xi_2-1)\ln\frac{Q^2}
{q_\perp^2}\right] \ ,\label{dfunc}
\end{eqnarray}
where the ``plus''-prescription is defined in the standard way
through
\begin{equation}
\int_x^1 dz \frac{f(z)}{(1-z)_+}= \int_x^1 dz
\frac{f(z)-f(1)}{1-z}+f(1)\ln(1-x) \ ,
\end{equation}
for any suitably regular function $f$.

After applying Eq.~(\ref{dfunc}) to the results presented in the
previous subsections, we find for the small-$q_\perp$ behavior of
the unpolarized cross section for $q+\bar q\to \gamma^*g$:
\begin{eqnarray}
\frac{d^4\sigma^{q\bar q\to
\gamma^*g}}{dQ^2dyd^2q_\perp}&=&\sigma_0\frac{\alpha_s}{2\pi^2}
C_F\frac{1}{q_\perp^2}\int\frac{dx}{x} \frac{dx'}{x'}\sum_q
e_q^2q(x)\bar q(x')
\left[\frac{1+\xi_1^2}{(1-\xi_1)_+}\delta(\xi_2-1)\right.\nonumber\\
&&\left.+ \frac{1+\xi_2^2}{(1-\xi_2)_+}\delta(\xi_1-1)+
2\delta(\xi_1-1)\delta(\xi_2-1)\ln\frac{Q^2}{q_\perp^2}\right] \
.\label{qqbunp}
\end{eqnarray}
For $q+g\to \gamma^*q$ scattering, we obtain
\begin{eqnarray}
\frac{d^4\sigma^{qg\to \gamma^*q}}{dQ^2dyd^2q_\perp}&=&\sigma_0
\frac{\alpha_s}{4\pi^2} \frac{1}{q_\perp^2} \int\frac{dx}{x}
\frac{dx'}{x'}\sum_q e_q^2 q(x)g(x')\delta(1-\xi_1) \left[\xi_2^2+
(1-\xi_2)^2\right] \ . \label{qgunp}
\end{eqnarray}
These results are well-known in the literature \cite{ColSopSte85}.

We now make similar expansions for the single-spin cross sections.
As we anticipated, at low transverse momentum, the soft- and
hard-pole contributions overlap: at $q_\perp =  0$, the
twist-three quark-gluon correlation function associated with the
hard-pole contributions, $T_F(x-\bar x_g,x)$, becomes
$T_F(x-x(1-\xi_1)/\xi_2,x)$, which at $\xi_1=1$ is identical to
the correlation function $T_F(x,x)$ that accompanies the soft
poles.

For the $q\bar q$ channel, we find:
\begin{eqnarray}
\frac{d^4\Delta\sigma^{q\bar q\to \gamma^*g}(S_\perp)}
     {dQ^2dyd^2q_\perp}
&=&\sigma_0\, \epsilon^{\alpha\beta}{S}_{\perp\alpha}\,
\frac{q_{\perp\beta}}{(q_\perp^2)^2}\,
\frac{\alpha_s}{2\pi^2}\int \frac{dx}{x}\frac{dx'}{x'} \bar q(x')
\left\{\delta(\xi_2-1)A+\delta(\xi_1-1)B\right\} \, , \label{e27}
\end{eqnarray}
where
\begin{eqnarray}
A&=& \frac{1}{2N_C} \left\{ \left[x\frac{\partial}{\partial
x}T_F(x,x)\right](1+\xi_1^2)
      +T_F(x,x-\widehat{x}_g)\frac{1+\xi_1}{(1-\xi_1)_+}\nonumber\right.\\
 &&\left.     +T_F(x,x)\frac{(1-\xi_1)^2(2\xi_1+1)-2}{(1-\xi_1)_+}\right\}
+C_F T_F(x,x-\widehat{x}_g)\frac{1+\xi_1}{(1-\xi_1)_+}\ ,
\label{qta}\\
B&=& C_F T_F(x,x)\left[\frac{1+\xi_2^2}{(1-\xi_2)_+}
+2\delta(\xi_2-1)\ln\frac{Q^2}{q_\perp^2}\right] \ , \label{qtb}
\end{eqnarray}
with $\widehat{x}_g\equiv(1-\xi_1)x$.

For the $qg$ channel, the contribution involving the derivative
$dT_F(x,x)/dx$ turns out to be of higher order in $q_\perp/Q$. The
full non-derivative term becomes
\begin{eqnarray}
\hspace*{-3mm} \frac{d^4\Delta\sigma^{qg\to \gamma^*q}(S_\perp)}
     {dQ^2dyd^2q_\perp}
&=& \sigma_0\, \epsilon^{\alpha\beta}{S}_{\perp\alpha}\,
    \frac{q_{\perp\beta}}{(q_\perp^2)^2}
\sum_qe_q^2T_F(z_1,z_1)
\nonumber \\
&\ & \times \frac{\alpha_s}{2\pi^2}T_R\int \frac{dx_2}{x_2}
g(x_2)\left[\xi_2^2+(1-\xi_2)^2\right]. \label{e26}
\end{eqnarray}

In the following section, we will reproduce the above results from
a QCD factorization in terms of transverse-momentum-dependent parton
distributions.

\section{Transverse-momentum-dependent QCD Factorization}

When $q_\perp\ll Q$, the Drell-Yan cross section in leading order
in $q_\perp/Q$ may be calculated from a QCD factorization theorem
involving transverse-momentum-dependent (TMD) parton distributions
\cite{ColSopSte85,JiMaYu04}. Therefore, one expects that the
results given in the previous section can be reproduced by this
alternative approach. Indeed, for the unpolarized cross section
this is well established~\cite{ColSopSte85,Laenen:2000ij}.

\subsection{Transverse-Momentum-Dependent Parton Distributions}

Transverse-momentum-dependent parton distributions were first
introduced by Collins and Soper~\cite{ColSop81}. They provide more
information about the structure of the nucleon than is contained
in the usual Feynman parton distributions. Based on rigorous
factorization theorems \cite{ColSop81,JiMaYu04,ColMet04}, these
distributions may be extracted from various processes that are
characterized by a large momentum scale and a much smaller,
measured, transverse momentum, such as semi-inclusive DIS and the
Drell-Yan process. In this paper, we follow a definition of the
TMD distributions in Feynman gauge with explicit gauge links
\cite{Col02}. We avoid the axial gauge because of the
complications presented by additional gauge links at space-time
infinity \cite{BelJiYua02}.

The TMD quark distributions of a polarized proton may be defined
through the following matrix:
\begin{eqnarray}
    {\cal M}^{\alpha\beta} &=&   P^+\int
        \frac{d\xi^-}{2\pi}e^{-ix\xi^-P^+}\int
        \frac{d^2b_\perp}{(2\pi)^2} e^{i\vec{b}_\perp\cdot
        \vec{k}_\perp} \, \left\langle
PS\left|\overline{\mit \Psi}_v^\beta(\xi^-,0,\vec{b}_\perp)
        {\mit \Psi}_v^\alpha(0)\right|PS\right\rangle\ ,
\end{eqnarray}
where the vector $P=(P^+,0^-,0_\perp)$ is along the momentum
direction of the proton, $S$ is the polarization vector, and ${\mit
\Psi}_v(\xi)$ is defined as
\begin{equation}
{\mit \Psi}_v(\xi) \equiv {\cal L}_{v}(-\infty;\xi)\psi(\xi)\ ,
\end{equation}
with the gauge link
\begin{equation}
 {\cal L}_{v}(-\infty;\xi) \equiv \exp\left(-ig\int^{-\infty}_0 d\lambda
\, v\cdot A(\lambda v +\xi)\right) \ .\label{glink}
\end{equation}
This gauge link goes to $-\infty$, indicating that we adopt the
definition for the TMD quark distributions for the {\it Drell-Yan
process} \cite{BroHwaSch02,Col02,BelJiYua02}. In a coordinate
system where $P^+\gg P^-$, the vector $v$ in the above equations
is taken to have $v^-\gg v^+, v_\perp=0$, with a nonzero $v^+$
component that regulates the light-cone singularities. The
leading-power expansion of the density matrix ${\cal M}$ contains
eight quark distributions \cite{MulTanBoe}. Here, we are only
interested in the TMD distributions for (un)polarized quarks and
antiquarks in the polarized proton, in particular in the Sivers
functions. The physics of the Sivers functions may be viewed as
follows. Consider a transversely polarized proton with large
momentum $P$. The distribution of quarks with longitudinal and
transverse momenta $xP$ and $\vec{k}_\perp$ can have a dependence
on the orientation of $\vec{k}_\perp$ relative to the proton's
polarization vector $\vec{S}_\perp$. Keeping only the unpolarized
quark distribution and the Sivers function, we have the following
expansion for the matrix ${\cal M}$:
\begin{equation}
{\cal M} = \frac{1}{2}
\left[q(x,k_\perp)\gamma_\mu P^\mu 
    + \frac{1}{M_P}\, q_T(x,k_\perp)\,
\epsilon_{\mu\nu\alpha\beta}\gamma^\mu P^\nu k^\alpha S^\beta +
\ldots\right] \label{matrixexp}
\end{equation}
where $q(x,k_\perp)$ is the TMD distribution in an unpolarized
proton, $q_T(x, k_\perp)$ is the Sivers function, and $M_P$ is a
hadron mass, used to normalize $q(x,k_\perp)$ and $q_T(x,k_\perp)$ to
the same mass dimension.

\begin{figure}[t]
\begin{center}
\includegraphics[height=10.0cm]{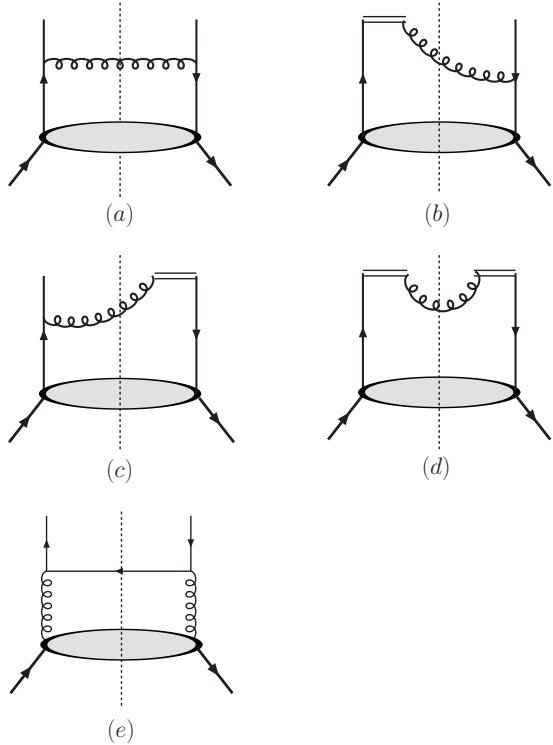}
\end{center}
\vskip -0.7cm \caption{\it Feynman diagrams contributing to the
spin-independent quark distribution at large transverse momentum
$k_\perp \gg \Lambda_{\rm QCD}$, arising from the integrated
quark~(a-d) and gluon~(e) distributions. Double lines denote
Eikonal lines.}
\end{figure}
When the transverse momentum $k_\perp$ is low, say, of order of
$\Lambda_{\rm QCD}$, the TMD parton distributions are entirely
non-perturbative objects. Thus, the only way of treating them is
to parameterize them in a suitable way, and to fit them to data.
However, for much larger transverse momentum,
$k_\perp\gg\Lambda_{\rm QCD}$, the $k_\perp$-dependence of the TMD
parton distributions can be calculated within perturbative QCD,
because a hard gluon has to be radiated in order to generate the
$k_\perp$. Through the pQCD calculation, the TMD parton
distributions may be related to the $k_\perp$-integrated parton
distributions (or, in case of the Sivers functions, to the
twist-three quark-gluon correlation function, as we shall show
below), multiplied by calculable hard coefficients. For example,
the TMD quark distribution for an unpolarized proton at large
$k_\perp$ will receive contributions from the $k_\perp$-integrated
quark and gluon distributions. The relevant Feynman diagrams are
shown in Fig.~8, and the results for the first-order term are
(see, for example, \cite{JiMaYu04,JiMaYu05}):
\begin{eqnarray}
q(z,k_\perp)&=&
\frac{\alpha_s}{2\pi^2}\frac{1}{\vec{k}_\perp^2}C_F\int\frac{dx}{x}
q(x) \left[\frac{1+\xi^2}{(1-\xi)_+}+\delta(\xi-1)
\left(\ln\frac{z^2\zeta^2}{\vec{k}_\perp^2}-1\right)\right]\nonumber\\
&&+
\frac{\alpha_s}{2\pi^2}\frac{1}{\vec{k}_\perp^2}T_R\int\frac{dx}{x}g(x)
\left[\xi^2+(1-\xi)^2\right]\ , \label{qktunp}
\end{eqnarray}
where $q(x)$ and $g(x)$ are the integrated quark and gluon
distributions, $\xi=z/x$, and $\zeta^2=(2v\cdot P)^2/v^2$. A
similar expression is obtained for the anti-quark TMD
distribution. We note that upon including virtual corrections, we
would find that the term $1/\vec{k}_\perp^2$ would be converted to
a term $1/(\vec{k}_\perp^2)_+$, and similarly for the logarithmic
term. This ``plus''-prescription would make the distribution
integrable at low $k_\perp$; for details, see for
example~\cite{aegm}.

\begin{figure}[t]
\begin{center}
\includegraphics[height=8.0cm]{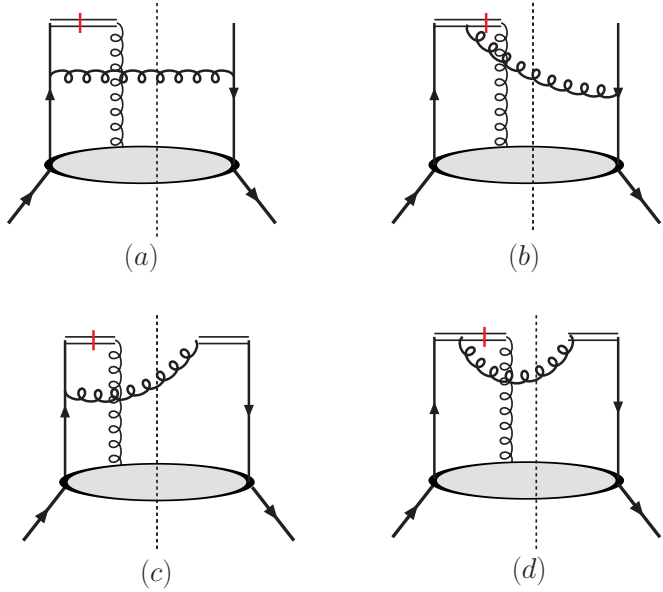}
\end{center}
\vskip -0.7cm \caption{\it Feynman diagrams contributing to the
Sivers functions at a large transverse momentum $k_\perp \gg
\Lambda_{\rm QCD}$, arising from the twist-three quark-gluon
correlation in Fig. 1: soft-pole contributions.}
\end{figure}

In the same spirit, the Sivers function at large $k_\perp$ can
also be calculated perturbatively. Because it is (naively)
time-reversal-odd, the only contribution comes from the
twist-three quark-gluon correlation, with the phase provided by
the hard part. This is very similar to the case of the single
transverse-spin dependent Drell-Yan cross section calculated in
Sec.~III. Also in the present calculation, we will have soft-pole
and hard-pole contributions, shown in Figs.~9 and 10,
respectively. Carrying out the calculations accordingly, we find
for the soft-pole contributions:
\begin{eqnarray}
q_T(z,k_\perp)|_{\rm soft-pole}
&=&\frac{\alpha_s}{4\pi^2}\frac{2M_P}{(\vec{k}_\perp^2)^2}
\frac{1}{2N_C}\int\frac{dx}{x}
\left\{\left[x\frac{\partial}{\partial
x'}T_F(x,x)\right]\left(1+\xi^2\right)\nonumber\right.\\
&+&\left. T_F(x,x)\left[\frac{(1-\xi)^2(2\xi+1)-2}{(1-\xi)_+}-
\delta(\xi-1)\left(\ln\frac{z^2\zeta^2}{k_\perp^2}-1\right)
\right]\right\} ,
\end{eqnarray}
where the $1/(k_\perp^2)^2$ behavior also follows from power
counting, and where $\xi$ and $\zeta$ as defined above after
Eq.~(\ref{qktunp}). We note that our choice of the direction of
the gauge link in Eq.~(\ref{glink}) is crucial since it determines
the sign of the Sivers
function~\cite{BroHwaSch02,Col02,BelJiYua02}. Only with the
correct choice (in this case, to $-\infty$) can the factorization
work out eventually. In case of semi-inclusive DIS a different
choice is necessary, and the resulting Sivers function will have
opposite sign.

The hard-pole contributions can be calculated similarly, and we
find
\begin{eqnarray}
q_T(z,k_\perp)|_{\rm hard-pole}
&=&\frac{\alpha_s}{4\pi^2}\frac{2M_P}{(\vec{k}_\perp^2)^2}
\left[C_F+\frac{1}{2N_C}\right]\int\frac{dx}{x} T_F(x,x-\widehat
x_g)\nonumber\\
&&\times \left[\frac{1+\xi}{(1-\xi)_+}+
\delta(\xi-1)\left(\ln\frac{z^2\zeta^2}{k_\perp^2}-1\right)
\right],
\end{eqnarray}
where $\widehat x_g=(1-\xi)x$. Adding the above soft-pole and
hard-pole contributions, we obtain the final result for the Sivers
function at large $k_\perp$:
\begin{eqnarray}
q_T(z_1,k_\perp)&=&\frac{\alpha_s}{4\pi^2}\frac{2M_P}
{(\vec{k}_\perp^2)^2}\int\frac{dx}{x} \left\{A+C_FT_F(x,x)
\delta(\xi_1-1)\left(\ln\frac{z_1^2\zeta_1^2}{\vec{k}_\perp^2}-
1\right)\right\} \ , \label{sivpert}
\end{eqnarray}
where $A$ has been given in Eq.~(\ref{qta}).

\begin{figure}[t]
\begin{center}
\includegraphics[height=7.0cm]{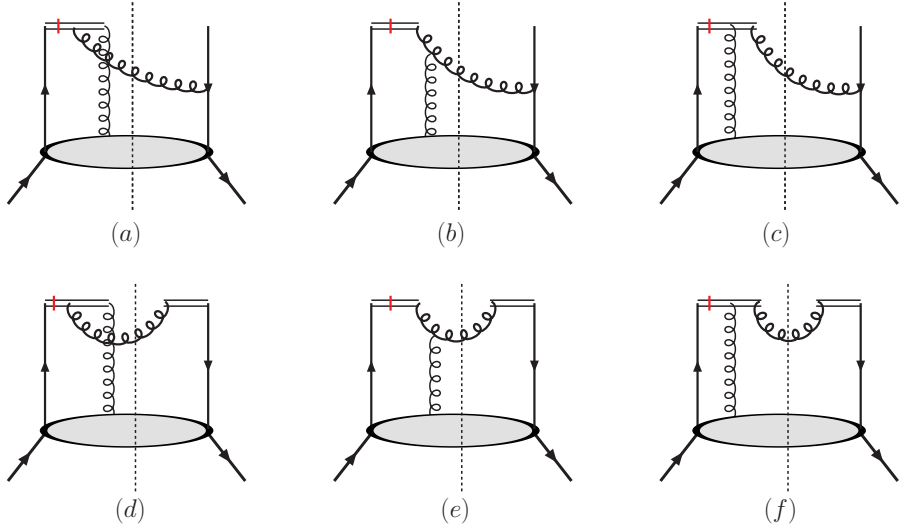}
\end{center}
\vskip -0.7cm \caption{\it Same as Fig.~9, but for the hard-pole
contributions.}
\end{figure}

\subsection{Unpolarized Cross Section}
The spin-independent cross section for the Drell-Yan process has
the following TMD factorization in the regime $q_\perp \ll
Q$~\cite{ColSopSte85,JiMaYu04}:
\begin{eqnarray}
\frac{d^4\sigma}{dQ^2dyd^2q_\perp}&=&\sigma_0\sum_q e_q^2\int
d^2\vec{k}_{1\perp}d^2\vec{k}_{2\perp}d^2
\vec{\lambda}_\perp\delta^{(2)}(\vec{k}_{1\perp}
+\vec{k}_{2\perp}+\vec{\lambda}_\perp-\vec{q}_\perp)
\nonumber\\
&&\times\, q(z_1,k_{1\perp},\zeta_1)\bar
q(z_2,k_{2\perp},\zeta_2)~H(Q^2)\left(S(\lambda_\perp,\rho)\right)^{-1}
\ , \label{qtfact}
\end{eqnarray}
where as before $q(z,k_\perp)$ and $\bar q(z,k_\perp)$ are the TMD
quark and antiquark distributions. $H$ is a hard factor and is
entirely perturbative. It is a function of $Q\gg q_\perp$ only.
The soft-factor $S$ is a vacuum matrix element of Wilson lines and
captures the effects of soft gluon radiation. Since the soft-gluon
contributions in the TMD distributions have not been subtracted,
the soft factor enters with inverse power. As mentioned earlier,
in order to regulate the light-cone singularities, we introduce
the off-light-cone vectors $v_1$ and $v_2$ for the TMD
distributions. We also define $\zeta_1^2= (2v_1\cdot P)^2/v_1^2$
and $\zeta_2^2= (2v_2\cdot P')^2/v_2^2$, and the soft-gluon
rapidity cut-off $\rho=\sqrt{(2v_1\cdot v_2)^2/v_1^2v_2^2}$. In a
special coordinate frame, $z_1^2\zeta_1^2=z_2^2\zeta_2^2=\rho Q^2$
\cite{JiMaYu04}. There are also explicit renormalization scale
dependences of the various factors in the factorization formula,
which have been omitted for simplicity.

In Eq.~(\ref{qktunp}) in the previous subsection, we have already
given the unpolarized TMD quark distribution, calculated
perturbatively for $k_\perp \gg \Lambda_{\rm QCD}$. Also the soft
factor may be calculated in this fashion. To leading order in
$\alpha_s$, one finds~\cite{JiMaYu04},
\begin{equation}
S(\lambda_\perp)=\delta^{(2)}(\vec{\lambda}_\perp) +
\frac{\alpha_s}{2\pi^2}\frac{1}{\vec{\lambda}_\perp^2}C_F\left(\ln
\rho^2-2\right) \ .\label{softpert}
\end{equation}
By adding virtual contributions, the soft function is normalized
in such a way that it obeys $\int d^2\vec{\lambda}_\perp
S(\lambda_\perp)=1$.

Inserting the perturbative TMD distribution~(\ref{qktunp}) and the
soft function~(\ref{softpert}) into the factorization
formula~(\ref{qtfact}), we find that indeed the unpolarized cross
section given by Eqs.~(\ref{qgunp}), (\ref{qqbunp}) is reproduced,
including the quark-gluon scattering piece. Here we use the above
normalization condition for the soft function, and the
normalization that the integration over the TMD distribution
yields the normal Feynman parton distributions,
\begin{eqnarray}
\int d^2\vec{k}_\perp q(z_1,k_\perp)= q(z_1)\ ,\quad \int
d^2\vec{k}_\perp \bar q(z_2,k_\perp)= \bar q(z_2)\ .
\end{eqnarray}

\subsection{Polarized Cross Section}

In~\cite{JiMaYu04} also a factorization formula for the
transverse-spin dependent Drell-Yan cross section at $q_\perp \ll
Q$ was established, which reads:
\begin{eqnarray}
\frac{d^4\Delta\sigma(S)}{dQ^2dyd^2q_\perp} &=&\sigma_0\,
\epsilon^{\alpha\beta}{S}_{\perp\alpha}\, q_{\perp\beta}
\frac{1}{M_P}\int
d^2\vec{k}_{1\perp}d^2\vec{k}_{2\perp}d^2\vec{\lambda}_\perp
~\frac{\vec{k}_{1\perp}\cdot
\vec{q}_\perp}{q_\perp^2}~\delta^{(2)}(\vec{k}_{1\perp}+
\vec{k}_{2\perp}+\vec{\lambda}_\perp-\vec{q}_\perp)
\nonumber\\
&&\times q_T(z_1,k_{1\perp},\zeta_1)~\bar
q(z_2,k_{2\perp},\zeta_2)~H(Q^2)\left(S(\lambda_\perp)\right)^{-1}
\ , \label{fac}
\end{eqnarray}
where $q_T(z,k_\perp)$ is the Sivers function defined in
Eq.~(\ref{matrixexp}), $\bar{q}(z,k_\perp)$ is again the
anti-quark TMD for the unpolarized proton, and $H$ and $S$ are the
hard and soft factors introduced earlier.

We have given the first-order perturbative result for the Sivers
function in Eq.~(\ref{sivpert}). As was shown in
\cite{BoeMulPij03} (see also~\cite{MaWa03,bacchetta}), its
$k_\perp^2$-moment is related to the twist-three quark-gluon
correlation function defined in Eq.~(\ref{Tdef}) of Sec.~II:
\begin{eqnarray}
\frac{1}{M_P} \int d^2\vec{k}_\perp \vec{k}_\perp^2
q_T(x,k_\perp)=T_F(x,x) \ .
\end{eqnarray}
For deep-inelastic scattering, this relation remains true, except
that the sign will be opposite. Inserting finally all TMD
functions into the factorization formula~(\ref{fac}), we reproduce
the differential Drell-Yan single-transverse-spin cross section
given at low transverse momentum in Eqs.~(\ref{e26}), (\ref{e27})
of Sec.~IV. This concludes our demonstration that the ETQS
mechanism and the approach based on TMD factorization produce
identical results in the kinematic regime where they overlap.

\section{Summary and Outlook}

In summary, we have studied the single-transverse-spin asymmetry
in Drell-Yan pair production at large and moderate transverse
momenta. At large transverse momenta $q_\perp\sim Q$, the
asymmetry is power-suppressed by $1/Q$. Therefore, the
collinear-factorized approach based on twist-three quark-gluon
correlations should provide the appropriate description. We have
used this approach to derive the single-spin asymmetry. We have
then expanded the result for moderate transverse momenta of the
lepton pair, $\Lambda_{\rm QCD}\ll q_\perp\ll Q$. In this regime,
one knows that a factorization theorem in terms of
transverse-momentum dependent parton distributions holds,
involving in particular the Sivers function. We have verified that
indeed a smooth transition  from the higher-twist mechanism to the
one based on TMD factorization occurs, with the two approaches
describing the same physics in the region of overlap. This unifies
the two approaches, and provides input for new strategies in the
analysis of experimental data for single-spin asymmetries.

In \cite{qiu}, it was noted that there are additional twist-three
quark-gluon correlation functions that may contribute to the SSAs
in hard processes. One of them is a spin-averaged and chiral-odd
correlation, defined by \cite{qiu}
\begin{equation}
T_F^{(\sigma)}(x_1,x_2)=
\int\frac{d\zeta^-d\eta^-}{8\pi}e^{ix_1P^+\eta^-}e^{i(x_2-x_1)P^+\zeta^-}
\left\langle P|\overline\psi(0)\sigma
^{+\alpha}gF^{+\alpha}(\zeta^-)\psi(\eta^-)|P\right\rangle \ .
\end{equation}
When combined with the quark transversity distribution for the
polarized proton, the above correlation in the unpolarized proton
can also contribute to the SSA for the Drell-Yan process at large
transverse momentum. However, after a straightforward calculation we
find that the contribution vanishes in the low transverse momentum limit
$q_\perp\ll Q$. This observation is consistent with the TMD
factorization at low transverse momentum \cite{JiMaYu04}, where only
the Sivers function contributes to the SSA for the Drell-Yan process
after integration over the lepton angles.

There are a number of directions in which our work could be
extended. First, an interesting topic will be to study the
single-spin asymmetry in semi-inclusive deep inelastic scattering
(SIDIS) $ep\to e \pi X$ \cite{yuji}, again at large and moderate
transverse momenta of the final-state hadron. Also here, a
connection between the twist-three mechanism and the TMD
factorization formalism can be established. The Sivers function
calculated perturbatively in this paper may be immediately applied
to SIDIS, after changing its sign. Although the calculation for
SIDIS should be straightforward following our derivations for the
Drell-Yan process in this paper, the details will differ. It will
be crucial to demonstrate that the two approaches are indeed also
consistent for the SIDIS process at $\Lambda_{\rm QCD}\ll
q_\perp\ll Q$, and to explicitly verify the process-dependence for
the single-transverse spin asymmetries in hard processes. It also
needs to be demonstrated that the Collins function
mechanism~\cite{collins} can be reproduced in the low-$q_\perp$
region. We will present a study of SIDIS in a future publication.

Second, the consistency between the twist-three and the TMD
factorization approaches should also lead the way to resumming the
large logarithms from higher-order soft gluon radiation. As shown in
Eq.(\ref{e27})-(\ref{qtb}), the first-order expressions contain a
large logarithm $\propto \ln Q^2/q_\perp^2$. This large logarithm
will be present in higher orders, with two additional powers of the
logarithm at every new order in perturbation. In order to have a
reliable theoretical calculation, these logarithms need to be
resummed. The resummation procedure may follow the classic example
in \cite{ColSopSte85} for the unpolarized Drell-Yan cross section.
The relevant Collins-Soper equation for the energy evolution of the
spin-dependent TMD quark distributions has already been derived in
\cite{IdiJiMaYu04}. The resummation effects will be particularly
important at small $q_\perp/Q$.

\section*{Acknowledgments}
We are grateful to G.~Sterman for useful discussions. X.~J. is
supported by the U. S. Department of Energy via grant
DE-FG02-93ER-40762 and by a grant from Chinese National Natural
Science Foundation (CNSF). J.~Q. is supported in part by the U. S.
Department of Energy under grant No. DE-FG02-87ER-40371. W.~V. and
F.~Y. are finally grateful to RIKEN, Brookhaven National
Laboratory and the U.S. Department of Energy (contract number
DE-AC02-98CH10886) for providing the facilities essential for the
completion of their work.

\end{document}